\documentclass[aps,prl,twocolumn]{revtex4-1}
\usepackage{graphicx}
\usepackage{dcolumn}
\usepackage{bm}
\usepackage{color}

\begin{document}

\title{Anionic depolymerization transition in IrTe$_{2}$}
\author{Yoon Seok Oh$^{1}$, J. J. Yang$^{2}$, Y. Horibe$^{1}$, and S.-W. Cheong$^{1,2,*}$}

\address{$^{1}$Rutgers Center for Emergent Materials and Department of Physics \& Astronomy, Piscataway, NJ 08854, USA\\
$^{2}$Laboratory for Pohang Emergent Materials and Department of Physics, Pohang University of Science and Technology, Pohang 790-784, Korea}

\date{\today}

\begin{abstract}
Selenium (Se) substitution drastically increases the transition temperature of iridium ditelluride (IrTe$_{2}$) to a diamagnetic superstructure from 278 K to 560 K. Transmission electron microscopy experiments revealed that this enhancement is accompanied by the evolution of non-sinusoidal structure modulations from $q = 1/5(10\bar{1})$- to $q = 1/6(10\bar{1})$-types. These comprehensive results are consistent with the concept of the destabilization of polymeric Te-Te bonds at the transition, the temperature of which is increased by chemical and hydrostatic pressure and by the substitution of Te with the more electronegative Se. This temperature-induced depolymerization transition in IrTe$_{2}$ is unique in crystalline inorganic solids.
\end{abstract}


\maketitle

A polymer is a large molecule made up of repeating chemical units connected by covalent bonds \cite{01HRAllock}. Many polymers are thermally unstable and decompose at high temperatures ($T$). Other polymers undergo phase transitions upon changes in temperature. Rubbery or flexible thermoplastics can transform to glassy or crystalline states below specific temperatures without losing their polymeric nature. Depolymerization-polymerization transitions have also been observed in a number of polymeric materials. For example, rhombic sulfur with cyclic molecule rings (depolymerized S$_{8}$) becomes polymerized to form long polymerized sulfur chains above 140 $^{\circ}$C \cite{02RFBacon,03REPowell}, and this depolymerization-polymerization transition is thermally reversible. Fullerene (C$_{60}$) molecules forming an fcc structure with a quasi-free molecular rotation \cite{04PHeiney} can be polymerized to orthorhombic one-dimensional chains \cite{05RMoret} or rhombohedral two-dimensional lattices \cite{06YIwasa} under external pressure. In addition, visible or ultraviolet light illumination can induce the polymerization of C$_{60}$ molecules \cite{07AMRao}.

Layered chalcogenides, composed of stacking polyhedron layers with van der Waals (VDW) gaps, exhibit rich quasi-low-dimensional physical properties such as superconductivity (in FeSe \cite{08FCHsu,09SMedvedev}), topological insulating behavior (in Bi$_{2}$Se$_{3}$ \cite{10MHasan,11KLQi,12YZhang}), and a high mobility in field-effect-transistor structures (in MoS$_{2}$: 200 cm$^{2}$V$^{-1}$s$^{-1}$ \cite{13VPodzorov}; in WSe$_{2}$: 500 cm$^{2}$V$^{-1}$s$^{-1}$ \cite{14BRadisavljevic}). Transition metal dichalcogenides (such as 1T-TaS$_{2}$, 1T-TaSe$_{2}$, and 1T-TiSe$_{2}$, forming layered CdI$_{2}$ structures (space group: $P\bar{3}m1$)) show Charge-Density-Wave (CDW) states accompanied by structural modulations below their transition temperatures ($T_{\rm C}$s). For 1T-TaS$_{2}$ and 1T-TiSe$_{2}$, applied pressure and chemical doping suppress the CDW state and induce superconductivity \cite{15BSipos,16AFKusmartseva}, attributed to competition between CDW and superconductivity. IrTe$_{2}$ also crystallizes in the CdI$_{2}$ structure. A sudden increase in resistivity below $\sim$260 K accompanying structural modulation and diamagnetism was interpreted in terms of CDW instability. Similarly to the behavior of 1T-TaS$_{2}$ and 1T-TiSe$_{2}$, superconductivity in Pd-intercalated and doped IrTe$_{2}$ was also understood in terms of the competition between CDW and superconductivity \cite{17AFFang}. However, a recent optical spectroscopy study suggested that the electronic/structural transition in IrTe$_{2}$ is driven by a reduction in the kinetic energy of electrons due to Te $5p$ band splitting below the structural transition temperature \cite{17AFFang}. Studies using x-ray photoemission spectroscopy \cite{18DOotsuki} and angle-resolved photoemission spectroscopy (ARPES) \cite{19DOotsuki} have indicated the importance of the orbital degeneracy of Ir $5d$ and/or Te $5p$ for the transition. Thus, the origin of the electronic/structural transition below $\sim$260 K remains unclear.

\begin{figure}
\begin{center}
\includegraphics[width=0.48\textwidth]{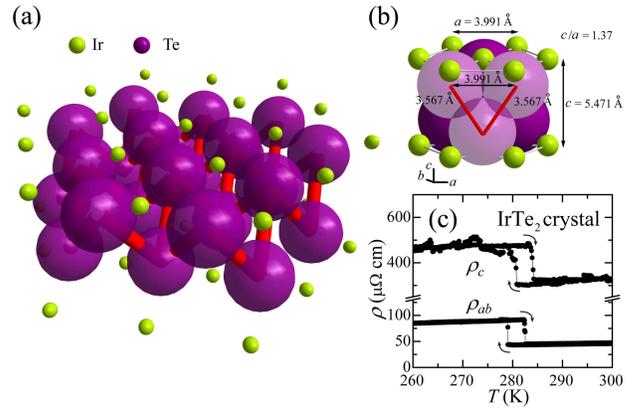}
\end{center}
\caption{(color online) (a) Schematic perspective view of the polymeric network of Te-Te bonds. (b) Crystal structure of IrTe$_{2}$ showing Te-Te bond lengths and the relative sizes of the Ir$^{4+}$ and Te$^{2-}$ ions (the radii of the Ir$^{4+}$ and Te$^{2-}$ ions with 6 coordination sites are 0.62 {\AA} and 2.2 {\AA}, respectively). (c) Temperature dependence of the resistivity of an IrTe$_{2}$ single crystal along the $c$-axis ($\rho_{c}$) and in the $ab$-plane ($\rho_{ab}$).}
\label{fig1}
\end{figure}

This letter provides evidence that the electronic/structural transition in IrTe$_{2}$ is a phase transition involving the depolymerization-polymerization of anionic Te bonds. Short Te-Te bonds between adjacent Te layers in the normal state of IrTe$_{2}$ have already been suggested to result in three-dimensional polymeric networks with multiple covalent Te-Te bonds (Figure 1(a)) \cite{20SJobic,21ECanadell}. Formation of the polymeric Te-bond networks in the polymerized state is associated with a fractional ionic character of Te (Te$^{1.5-}$) and with the destabilization of the highly oxidized state of Ir, resulting in effective Ir$^{3+}$ valence states \cite{22SJobic}. Consistently, the Te-Te distance (3.567 {\AA}) between neighboring Te-Te layers is $\sim$10 $\%$ shorter than that in each Te layer \cite{21ECanadell} (Figure 1(b)). As a result, IrTe$_{2}$ has a $c/a$ ratio of 1.37, significantly smaller than the $c/a$ ratio of 1.6-1.8 for the typical hexagonal close-packed CdI$_{2}$ structure with VDW gaps \cite{23SJobic}. For example, in HfTe$_{2}$ (1T-TiSe$_{2}$), where Hf (Ti) ions are in a stable, highly oxidized state of $4+$, VDW gaps are allowed between the Te-Te (Se-Se) layers, and thus HfTe$_{2}$ (1T-TiSe$_{2}$) exhibits a large $c/a$ ratio of 1.68 (1.70) \cite{22SJobic,23SJobic}. Combined with the results of an earlier x-ray spectroscopy study \cite{18DOotsuki}, our comprehensive experimental results on IrTe$_{2}$ with Se substitution and hydrostatic pressure strongly suggest that below $\sim$260 K, the covalent Te-Te bonds weaken and the polymeric Te-bond network becomes depolymerized. This depolymerization of Te-Te bonds is associated with an increase in the ionic character of Te$^{1.5-\delta/2-}$ and a mixed valence state of Ir$^{3+\delta+}$ in the diamagnetic superstructure, where $\delta$ indicates the change in the Ir valence state.

IrTe$_{2-x}$Se$_{x}$ specimens were prepared in polycrystalline and single crystalline forms. For polycrystalline IrTe$_{2-x}$Se$_{x}$, Ir, Te, and Se elements were mixed stoichiometrically, ground, pelletized, and synthesized in vacuum-sealed quartz ampules. X-ray diffraction (XRD) experiments were performed using a Rigaku D/Max-RB x-ray diffractometer with Cu $K_{\alpha}$ radiation. Single crystalline samples were grown from the Te metal flux. The Se compositions of IrTe$_{2-x}$Se$_{x}$ single crystals were estimated by comparing their $T_{\rm C}$s with those of polycrystalline IrTe$_{2-x}$Se$_{x}$. Magnetization and electrical resistivity were measured up to 400 K using the Quantum Design MPMS-XL7 and PPMS-9. High-$T$ transport properties were measured in a tube furnace using the DS340 function generator and an SR510 lock-in amplifier, and hydrostatic pressure experiments were performed using Easylab Pcell30. Samples for transmission electron microscopy (TEM) studies were prepared by cleaving and Ar ion milling. TEM experiments were carried out with JEOL-2010F and JEOL-2000FX transmission electron microscopes equipped with a low-$T$ sample stage and a room-$T$ double-tilt sample stage.

\begin{figure}
\begin{center}
\includegraphics[width=0.48\textwidth]{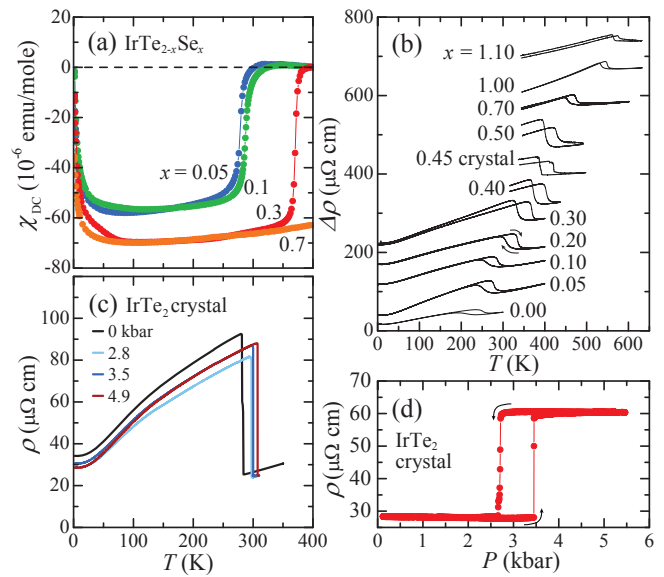}
\end{center}
\caption{(color online) Temperature dependence of (a) the magnetic susceptibility of polycrystalline IrTe$_{2-x}$Se$_{x}$ in $\mu_{0}H$=1 T upon warming, and (b) the resistivity of polycrystalline IrTe$_{2-x}$Se$_{x}$ and a single crystal. $\rho(T)$ curves for $x$ values other than $x$=0 are shifted by arbitrary constants for clarity. (c) The temperature dependence of the resistivity of an IrTe$_{2}$ single crystal upon warming under various hydrostatic pressures. (d) Pressure dependence of the resistivity of an IrTe$_{2}$ single crystal at 300 K.}
\label{fig2}
\end{figure}

Figure 2(a) exhibits the Se substitution effect on $\chi_{\rm DC}(T)$ of polycrystalline IrTe$_{2-x}$Se$_{x}$, measured upon warming. As Se concentration increases, the transition to the low-$T$ diamagnetic state increases significantly from 278 K ($x$=0.05) to 288 K ($x$=0.1) and 370 K ($x$=0.3). At $x$=0.7, the material only shows diamagnetic susceptibility up to 400 K. This $T_{\rm C}$ enhancement is more evident in Fig. 2(b), which shows the $T$ dependence of electric resistivity ($\rho(T)$) of polycrystalline specimens up to $x$=1.1. A maximum $T_{\rm C}$ of $\sim$560 K was observed for IrTe$_{0.9}$Se$_{1.1}$, corresponding to the chemical solubility limit of Se. A single crystal of IrTe$_{2}$ undergoes its transition at 283 K (279 K) upon warming (cooling), as shown in Fig. 1(c). As shown in Fig. 2(c), applying hydrostatic pressure ($P$) to an IrTe$_{2}$ single crystal also increases $T_{\rm C}$. The sharp transition in ambient pressure increases monotonically with increasing hydrostatic pressure. At 300 K, hydrostatic pressure consistently induces the transition at 3.5 kbar (2.7 kbar) upon increasing (decreasing) pressure.

\begin{figure}
\begin{center}
\includegraphics[width=0.48\textwidth]{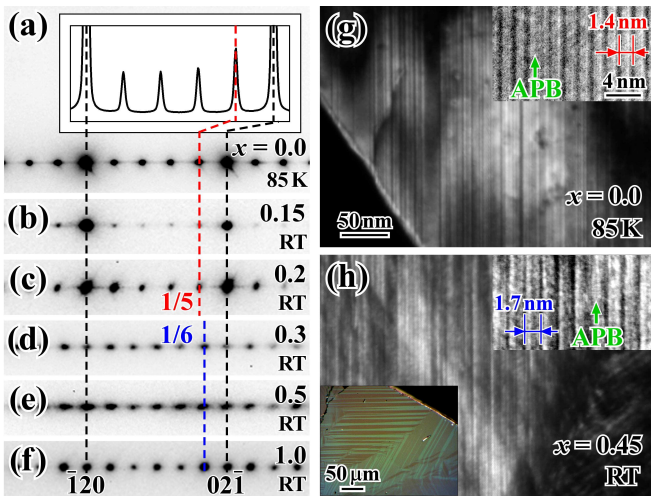}
\end{center}
\caption{(color online) (a)-(f) Electron diffraction patterns between the $\bar{1}20$ and $02\bar{1}$ fundamental spots for IrTe$_{2-x}$Se$_{x}$. The intensity profile between the $\bar{1}20$ and $02\bar{1}$ fundamental peaks of $x$=0.0 is shown in the inset. Dark-field images in (g) $x$=0.0 at $\sim$85 K, and (h) $x$=0.45 at room temperature. The upper right insets of (g) and (h) are high-resolution TEM images for $x$=0.0 at $\sim$85 K and $x$=0.45 at room temperature, respectively. The (g) and (h) insets show the superlattice modulations with periodicities of 1.4 nm and 1.7 nm associated with the $1/5(10\bar{1})$ and $1/6(10\bar{1})$ superlattice peaks, respectively, as well as antiphase boundaries (APB) due to phase shifts of the superlattice modulations. The lower left inset of (h) displays a polarized optical microscope image of twin domains of $x$=0.45.}
\label{fig3}
\end{figure}

Note that the magnitude of diamagnetic susceptibility suddenly increases from approximately $-60\times10^{-6}$ emu/mole ($x$=0.0 \cite{24JYang}, 0.05 and 0.1) to $-70\times10^{-6}$ emu/mole ($x$=0.3 and 0.7). The estimated core diamagnetism is approximately $-170\times10^{-6}$ emu/mole, and the difference between the core diamagnetism and the observed diamagnetic signals corresponds to the contribution of Pauli paramagnetism from itinerant electrons. This sudden change in Pauli paramagnetism reflects an abrupt change in the electronic structure of IrTe$_{2-x}$Se$_{x}$ with respect to Se doping. Indeed, TEM below $T_{\rm C}$ reveals the existence of two distinct modulated structures, characterized by the presence of two distinct superlattice peaks. Figures 3(a)-3(f) show electron diffraction patterns around the $\bar{1}20$ and $02\bar{1}$ fundamental spots for $x$=0.0, 0.15, 0.2, 0.3, 0.5, and 1.0, respectively. Figure 3(a) presents data for pure IrTe$_{2}$ obtained at $\sim$85 K, while Figs. 3(b)-3(f) present data recorded at room $T$. The $1/5(10\bar{1})$-type superlattice spots observed at $x$=0.0 (Fig. 3(a)) remain intact for $x$=0.15 and 0.2 (Fig. 3(b) and 3(c)) but change suddenly to the $1/6(10\bar{1})$-type at $x\geq0.3$ (Figure 3(d)-3(f)). The $1/5(10\bar{1})$ and $1/6(10\bar{1})$ superlattice peaks correspond to 1.4 nm and 1.7 nm spacings in real space, respectively. In both cases, we have frequently observed fine antiphase boundaries within a large strain-relieving twin domain in dark-field images obtained using a $02\bar{1}$ fundamental spot. Dark-field images of $x$=0.0 at $\sim$85 K and $x$=0.45 at room $T$ are shown in Fig. 3(g) and 3(h), respectively. Fine antiphase boundaries perpendicular to the superlattice modulation wave vectors are clearly visible in both Fig. 3(g) and 3(h). The typical spacing between antiphase boundaries is approximately 10 nm in both cases. High-resolution TEM images in the upper right insets of Fig. 3(g) and 3(h) clearly indicate that these antiphase boundaries are associated with phase shifts of the superlattice modulation waves across the boundaries. The intensity profile between the $\bar{1}20$ and $02\bar{1}$ fundamental peaks of $x$=0.0 (inset of Fig. 3(a)) reveals that all four $1/5(10\bar{1})$-type superlattice peaks exhibit similar intensities. A similar trend can be observed for the superlattice peaks of other compositions. The presence of high harmonics with strong intensities indicates that the superstructure modulations are highly non-sinusoidal and rather rectangular. This highly non-sinusoidal modulation is different from the typical sinusoidal modulation in CDW states, for example, in 1T-TaSe$_{2}$ and 1T-TaS$_{2}$ \cite{25JAWilson}. In addition, most CDW transitions in $MX_{2}$ are associated with three $q$ modulations, but the modulation in IrTe$_{2}$ occurs together with a $``$single $q$ and three domains$"$. Indeed, the three domains (or twins), corresponding to three $q$ values with a relative 120$^{\circ}$ in-plane angle, are visible in polarized optical microscope images (lower left inset in Fig. 3(h)) as well as in TEM images (see Fig. S1).

Figure 4(a) presents the $T$ versus $x$ phase diagram of IrTe$_{2-x}$Se$_{x}$, where the phase boundaries are determined from the results of TEM and $\rho(T)$ experiments with warming and cooling. The thermal hysteresis $({\it \Delta}T)$ of $T_{\rm C}$, estimated from warming and cooling $\rho(T)$ curves as a function of $x$ (see Fig. 4(b)), exhibits a sudden jump at $x$=0.2-0.3, coinciding with the abrupt change of superstructure from the $1/5(10\bar{1})$ to the $1/6(10\bar{1})$ type. Thus, although $T_{\rm C}$ appears to increase monotonically as a function of $x$ across the phase boundary of the $1/5(10\bar{1})$ and $1/6(10\bar{1})$ superstructures, ${\it \Delta}T$ accurately reflects the presence of the phase boundary.

XRD experiments at room $T$ are highly informative regarding the physical nature of the transition. As shown in Fig. 4(c), the $a$- and $c$-axis lattice constants for $x\leq0.1$ decrease linearly with increasing $x$ ($i.e.$, following Vegard's law), consistent with the fact that Se ions are smaller than Te ions. However, a sudden increase in $``c"$ (along with a smaller increase in $``a"$) occurs at $x$=0.1-0.15, corresponding to the phase boundary (purple vertical line) between the high-$T$ polymerized state and the $1/5(10\bar{1})$ superstructure. Thus, the dominant structural effect at the transition to the diamagnetic superstructure is the sudden increase in $``c"$, which reflects the sudden weakening of interlayer coupling. This increase in $``a"$ and $``c"$ gradually lessens for values of $x$ beyond the phase boundary (green vertical line) between the $1/5(10\bar{1})$ and $1/6(10\bar{1})$ superstructures. The structural evolution over $x$ seems better reflected in a plot of the $c/a$ ratio vs. $x$. The $c/a$ ratio increases drastically at the phase boundary between the high-$T$ polymerized state and the diamagnetic $1/5(10\bar{1})$ superstructure, suggesting a weakening of the polymerized Te bonds through the first-order phase transition. As the $c/a$ ratio (1.39-1.4) is still significantly smaller than that (1.6-1.8) of the true VDW systems, the depolymerization in the diamagnetic superstructure is likely still partial. In fact, the modulated superstructure may result from an ordered arrangement of polymerized and depolymerized Te bonds. We emphasize that the most drastic observation in our experimental results is the large increase in $T_{\rm C}$ resulting from the substitution of Se ions, which not only have a smaller ionic radius but are also more electronegative than Te ions. These electronegative Se ions should destabilize covalent anionic bonding, thus leading to the drastic increase in $T_{\rm C}$.

\begin{figure}
\begin{center}
\includegraphics[width=0.48\textwidth]{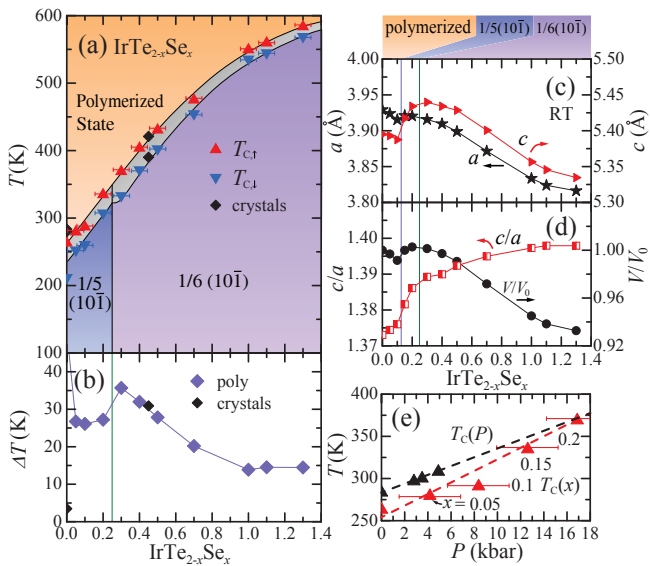}
\end{center}
\caption{(color online) (a) Temperature ($T$) versus Se concentration ($x$) phase diagram of IrTe$_{2-x}$Se$_{x}$ constructed with $T_{\rm C}$ values obtained from $T$-dependent resistivity measurements under warming (triangle) and cooling (inverted triangle). The diamond symbols indicate the $T_{\rm C}$ values of IrTe$_{2-x}$Se$_{x}$ single crystals ($x$=0 and 0.45). (b) Width of the thermal hysteresis (${\it \Delta}T$) as a function of $x$ in IrTe$_{2-x}$Se$_{x}$. The green guide line depicts the phase boundary between the $1/5(10\bar{1})$ and $1/6(10\bar{1})$ superstructures. Se-doping dependence of (c) the $a$- and $c$-axis lattice constants, (d) the $c/a$ ratio and the normalized volume determined from x-ray diffraction results. The purple and green lines indicate the phase boundaries of the paramagnetic to $1/5(10\bar{1})$ transition and the $1/5(10\bar{1})$ to $1/6(10\bar{1})$ transitions, respectively. (e) Hydrostatic pressure and calculated chemical pressure dependence of $T_{\rm C}$ of IrTe$_{2}$.}
\label{fig4}
\end{figure}

Furthermore, the increase in the number of depolymerized Te bonds with increasing Se substitution is also consistent with the decrease in the magnitude of the modulating wave vector from a 1/5 to 1/6 type. We also note that this scenario is consistent with the weakening of anionic polymeric Te-Te bonds with Se doping observed in Pt(Te,Se)$_{2}$ \cite{26GKliche}. Depolymerization reorganizes the ionic characteristics of Te anions, and reorganized Te$^{1.5-\delta/2-}$ ions lead to a mixed valence state of Ir cations such as Ir$^{3+\delta+}$. In fact, evidence of mixed valence Ir ions in the low-$T$ diamagnetic superstructure was observed in a previous study using x-ray spectroscopy \cite{18DOotsuki}. Note that the non-sinusoidal structural modulations in the TEM experiments here appear to be consistent with the charge ordering of the mixed valence states of Ir ions.

The increase in $T_{\rm C}$ with increasing hydrostatic pressure is summarized in Fig. 4(e). Thus, hydrostatic pressure also destabilizes the polymerized state, which seems to be counter-intuitive. However, the application of hydrostatic pressure to PtTe$_{2}$ \cite{27CSoulard} with anionic polymeric Te-Te bonds increases the ratio of interlayer to intralayer Te-Te distance. Thus, this increase in the interlayer/intralayer ratio with hydrostatic pressure likely stabilizes the low $T$ depolymerized state, leading to the increase in $T_{\rm C}$. The magnitude of the chemical pressure resulting from Se doping has been estimated using the initial slope of $a(x)$ and $c(x)$ for $0\leq x\leq 0.1$ and the bulk modulus \cite{29SJobic, 28FBirch}. Fig. 4(e) compares the effect of mechanical hydrostatic pressure with the effect of Se chemical pressure on $T_{\rm C}$. This qualitative agreement indicates that the effect of Se substitution also fits the role of an archetypal chemical pressure effect. However, the positive slope with respect to chemical pressure, $\partial T_{\rm C}/\partial P = +6.8$ K/kbar, is larger than that in response to hydrostatic pressure, $\partial T_{\rm C}/\partial P = +5.4$ K/kbar. Thus, the depolymerization of the polymeric Te bond appears to be further enhanced by the electronegative character of Se ions. Note that the increase in $T_{\rm C}$ in IrTe$_{2}$ with increasing chemical/hydrostatic pressure is in contrast with the effect of chemical/hydrostatic pressure on transition metal dichalcogenides (such as 1T-TaS$_{2}$ \cite{15BSipos}, 1T-TiSe$_{2}$ \cite{16AFKusmartseva}, and 1T-TiSe$_{2-x}$S$_{x}$ \cite{30MMay}), in which the CDW state is suppressed by pressure.

A number of our findings from the chemical/hydrostatic pressure, TEM, and XRD experiments are distinct from the typical behavior of CDW systems. A peak near $1/5(10\bar{1})$ in the charge susceptibility of IrTe$_{2}$ was interpreted as an indication for the presence of a nesting vector of Fermi surface and CDW instability \cite{24JYang}, but the charge susceptibility exhibited other features. Our results, first, reveal that the superlattice modulations are associated with a single $q$ and three domains and are highly non-sinusoidal, uncommon characteristics of typical CDW systems. Second, the drastic increase in $T_{\rm C}$ up to 560 K with the application of chemical and hydrostatic pressure to IrTe$_{2}$ has never been observed in any other CDW systems, which tend to have modest $T_{\rm C}$ values due to the small energy scale associated with Fermi surface instability. In addition, the sudden increase in the $c/a$ ratio below $T_{\rm C}$ is also unique compared with the behavior of CDW systems. These empirical observations, combined with the earlier proposal of charge modulations based on x-ray photoemission spectroscopy experiments \cite{18DOotsuki}, are consistent with our proposed scenario of a depolymerization-polymerization transition at $T_{\rm C}$ associated with the charge ordering of Ir$^{3+}$/Ir$^{4+}$ ions. We emphasize that the earlier study of Fermiology \cite{24JYang} did not take into account the lattice contributions such as phonon contribution, and we cannot rule out that the Fermi surface instability contributes to the depolymerization-polymerization transition in a secondary manner. Full theoretical investigation of the intriguing transition including the lattice contributions will be necessary to unveil the microscopic origin of the transition.

In conclusion, this study has explored the unique nature of the first-order phase transition at $\sim$260 K in IrTe$_{2}$, leading to diamagnetism, an increase in resistivity, and non-sinusoidal superlattice modulations of the $1/5(10\bar{1})$ type. The weak ionic nature of Te and the stability of Ir$^{3+}$ lead to a polymerized state with anionic polymeric networks of covalent Te-Te bonds in adjacent Te layers. The effective valence of Te in this polymerized state is $1.5-$. Below $\sim$260 K, the covalence of Te-Te bonds is partially lost, and the Te-Te networks are depolymerized. This leads to a reduced metallic character and the appearance of a superstructure. The effective valences of Ir and Te in the low $T$ depolymerized state are $(3+\delta)+$ and $(1.5+\delta/2)-$, respectively. This depolymerized state is found to be drastically stabilized by chemical pressure and hydrostatic pressure, as evidenced by increases in the $c/a$ ratio and $T_{\rm C}$ with hydrostatic pressure and Se substitution. Substitution of Te with Se increases $T_{\rm C}$ up to 560 K and leads to a distinct transformation of the superstructure from $1/5(10\bar{1})$ to $1/6(10\bar{1})$. This reversible depolymerization-polymerization transition appears to be unique among crystalline inorganic solids. These findings provide a new facet in the research of layered chalcogenides, materials that have continuously drawn the attention of the condensed matter physics community over the last several decades.

The work at Rutgers was supported by the National Science Foundation DMREF 1233349. The work at Postech was supported by the Max Planck POSTECH/KOREA Research Initiative Program $[\#2011-0031558]$ through the National Research Foundation of Korea funded by the Ministry of Education, Science and Technology.

\end{document}